# Coordination-Sensitive Nanoscale Analysis of Defect-Driven Phase Transformation in Si-Doped $(Al_XGa_{1-X})_2O_3$

Shaon Das[1], Jith Sarker[1], Christopher Chae[3], Lingyu Meng[2], Joel B. Varley[4], Hongping Zhao[2, 3], Jinwoo Hwang[3], and Baishakhi Mazumder[1*]

[1] *Department of Materials Design and Innovation, University at Buffalo-SUNY, Buffalo, NY 14260, USA*
[2] *Department of Electrical and Computer Engineering, The Ohio State University, Columbus, OH 43210, USA*
[3] *Department of Materials Science and Engineering, The Ohio State University, Columbus, OH 43210, USA*
[4]*Materials Science Division, Lawrence Livermore National Laboratory, Livermore, CA, USA*

* *Corresponding author: baishakh@buffalo.edu*

**ABSTRACT:**

Defect-driven phase instability critically influences the structural reliability of ultrawide bandgap oxides, yet direct nanoscale metrics linking local chemistry to structural transformation remain limited. Here, we introduce a coordination-sensitive atom probe tomography framework that quantitatively resolves reductions in local cation coordination and links them directly to defect-driven phase transformation. Using Si-doped β-$(Al_XGa_{1-X})_2O_3$ heterostructures with controlled Al composition (6–17%) and doping levels ($10^{17}$–$10^{20}$ cm$^{-3}$), we show that γ-phase inclusions emerge exclusively under the combined conditions of elevated Al content and heavy Si doping. Two-dimensional compositional mapping reveals pronounced lateral Al/Ga inhomogeneity in these regions, while nearest-neighbor and radial distribution analyses quantitatively resolve a significant reduction in first-shell Ga coordination, consistent with local cation deficiency. Correlative scanning transmission electron microscopy confirms that these coordination-depleted regions coincide spatially with γ-phase inclusions. Density functional theory further supports this mechanism, demonstrating that Al incorporation reduces monoclinic lattice stability and, in conjunction with donor-induced vacancy formation, facilitates vacancy-mediated cation rearrangement and coordination collapse. Together, these results establish coordination loss as a measurable nanoscale signature directly linked to defect-driven phase instability. This framework provides a generalizable approach for probing defect-driven phase instability in doped and alloyed ultrawide bandgap semiconductors.

Among the various polymorphs of gallium oxide ($Ga_2O_3$), the stable beta-phase (β-$Ga_2O_3$) and its alloys have attracted significant attention for next-generation power and RF, and optoelectronic applications.[1–4] The ultra-wide bandgap (UWBG) of β-$Ga_2O_3$ (4.6–4.9 eV) and high critical electric field (~8 MV/cm) make it promising for high-voltage and high-frequency electronic devices.[5–11] Recent advances in epitaxial growth have demonstrated β-$Ga_2O_3$ devices with multi-kilovolt breakdown voltages and improved transport performance.[12,13] However, the fixed bandgap of β-$Ga_2O_3$ limits bandgap engineering and heterostructure design. Alloying with aluminum to form $(Al_XGa_{1-X})_2O_3$, enables bandgap tunability for heterostructure engineering and deep-ultraviolet applications.[14–17] In parallel, controlled n-type doping using silicon is essential for achieving high carrier concentrations and device functionality.[18–21] However, increasing evidence suggests that the combined effects of alloying and heavy donor doping can destabilize the β-phase through defect-driven structural transformations. Understanding how local defect chemistry governs this phase instability remains a critical challenge for the reliable design of ultrawide bandgap oxide semiconductors.

Beyond carrier control, Si incorporation in $\beta$-$(Al_XGa_{1-X})_2O_3$ can significantly influence local defect chemistry and cation coordination, particularly at high doping levels. First-principles studies indicate that Al preferentially occupies octahedral sites in $\beta$-$(Al_XGa_{1-X})_2O_3$ under equilibrium conditions, although non-equilibrium thin-film growth can also promote Al incorporation on tetrahedral sites[22], while Si donors preferentially occupy tetrahedral sites and may compete with Al for cation site occupancy at elevated Al concentrations.[23–25] In heavily doped alloys, these effects promote local chemical heterogeneity and enhance the susceptibility of the β-phase lattice to defect-driven structural instability. In addition to modifying the local bonding environment, heavy donor doping promotes the formation of compensating cation vacancies, which act as dominant acceptor defects in n-type $\beta$-$(Al_XGa_{1-X})_2O_3$, as evidenced experimentally and theoretically.[22,26,27] These vacancies are not isolated point defects but can undergo structural reconstruction into energetically favorable split-vacancy configurations, accompanied by local cation displacement and lattice relaxation.[28–30] Together, these interactions indicate that dopant–alloy coupling can drive local structural instability through vacancy formation and cation redistribution.

A critical challenge in $\beta$-$(Al_XGa_{1-X})_2O_3$ is understanding how defect formation influences structural and electrical behavior. Defects introduced through doping or alloying can locally perturb the atomic structure and form defect complexes with diverse electronic characteristics.[23,26,30,31] Of particular concern is the γ-phase, a metastable defective spinel polymorph that locally disrupts the thermodynamically stable β-phase.[22,32–36] This phase is commonly associated with strain, defect accumulation, heavy doping, and local lattice distortions.[27,37–41] Experimental studies further show that β-to-γ transformation can be promoted by ion implantation, elevated temperatures, and impurities.[41–47] Recent computational studies suggest that this transformation is closely linked to vacancy-mediated cation rearrangements within the lattice. Increasing vacancy concentrations can drive redistribution between tetrahedral and octahedral coordination through split-vacancy configurations, producing a local cation arrangement consistent with the spinel-like γ-phase.[28,29,48] Computational studies further suggest that the oxygen framework remains largely preserved during the transformation, while disorder develops primarily within the cation sublattice.[49,50] The transformation is enabled by accessible cation migration pathways, with native cations exhibiting more facile reorientation than donor dopants under typical metal organic chemical vapor deposition (MOCVD) growth conditions.[28,29,51] Such γ-phase like inclusions can degrade device performance in $\beta$-$Ga_2O_3$ and $\beta$-$(Al_XGa_{1-X})_2O_3$ systems, underscoring the importance of understanding and controlling their formation. Despite these advances, a direct experimental framework to quantify nanoscale coordination changes and link them to defect-driven phase transformation remains lacking.

Here, we introduce a coordination-sensitive atom probe tomography (APT) framework to directly correlate γ-phase formation with local compositional and structural variations in Si-doped β-$(Al_XGa_{1-X})_2O_3$. By combining APT with high-resolution scanning transmission electron microscopy (STEM) and density functional theory (DFT), we establish a coordination-sensitive nanoscale approach that quantitatively links local coordination depletion with defect-driven structural instability. Our results reveal that phase instability emerges from the coupled effects of alloying and dopant-induced defect formation, which together promote vacancy-mediated cation redistribution and structural destabilization of the β-phase lattice. More broadly, this work establishes coordination loss as a measurable nanoscale signature of phase instability and provides a generalizable framework for understanding defect-driven structural transformations in doped and alloyed ultrawide bandgap materials.

## RESULTS AND DISCUSSION

The Si-doped $(Al_XGa_{1-X})_2O_3$ heterostructures investigated in this work consist of three doped layers with nominal Si concentrations of ~$10^{20}$, $10^{18}$, and $10^{17}$ $cm^{-3}$, grown for two Al compositions of x ≈ 0.17 and x ≈ 0.06, as illustrated in **Figure 1a**. Specifically, **Figure 1b and 1c** show the corresponding one-dimensional atomic concentration profiles for the x ≈ 0.17 and x ≈ 0.06 heterostructures, respectively, confirming controlled incorporation of Al, Ga, and Si with well-defined interfaces between layers. Minor variations observed near the interfaces are consistent with expected diffusion gradients during MOCVD growth and do not indicate significant intermixing. Within this structure, a pronounced lateral compositional variation emerges exclusively in the $(Al_{0.17}Ga_{0.83})_2O_3$ layer doped with ~$10^{20}$ $cm^{-3}$ Si (**Figure 2**), while layers with lower Si concentrations remain chemically uniform. This behavior is not observed in any other doped layer within the structure.

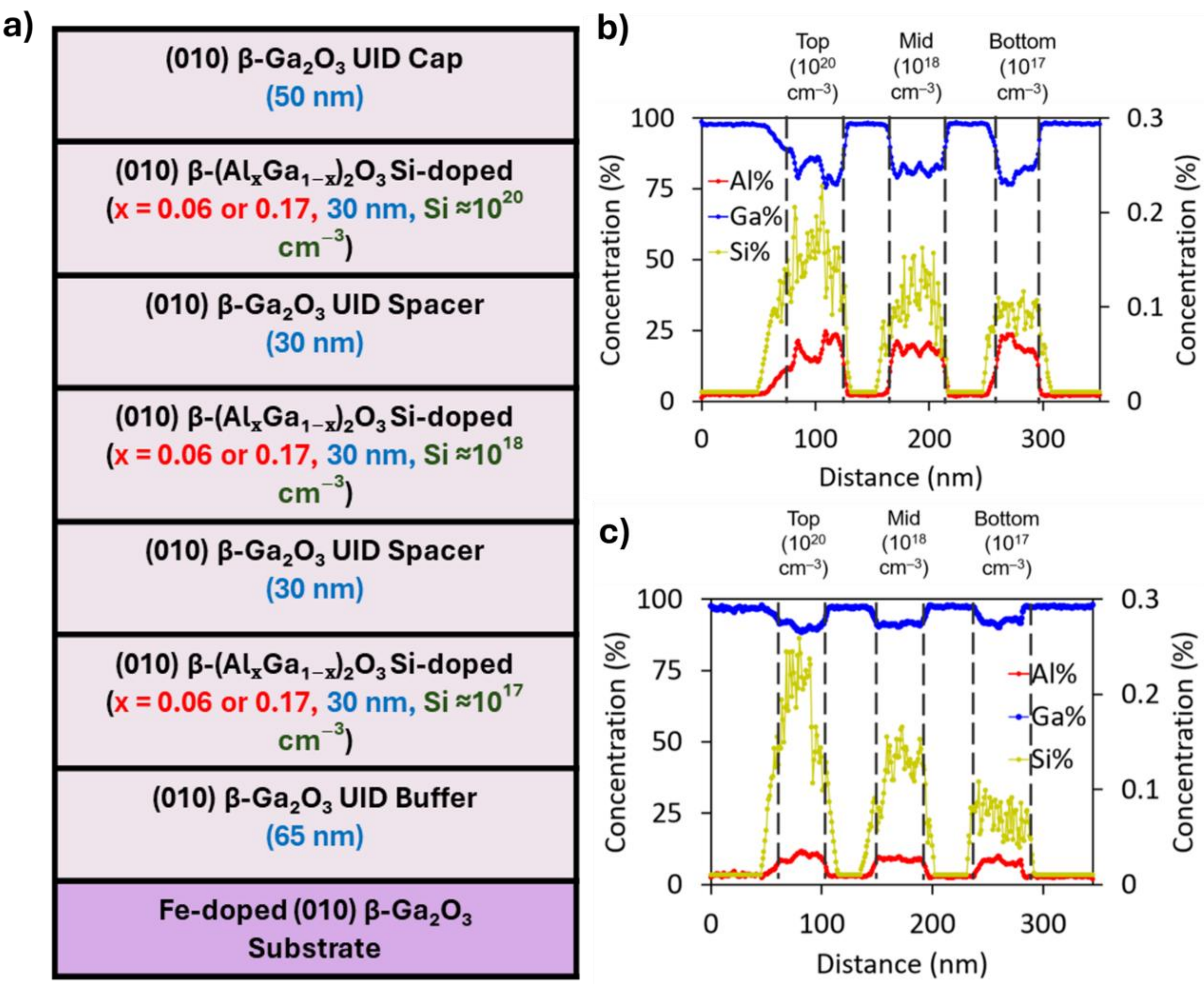

**Figure 1.** (a) Layer schematic of the Si-doped $(Al_XGa_{1-X})_2O_3$ heterostructure grown on a Fe-doped $\beta$-$Ga_2O_3$ substrate, incorporating three Si doping concentrations (~$10^{20}$, $10^{18}$, and $10^{17}$ $cm^{-3}$) across $\beta$-$(Al_XGa_{1-X})_2O_3$ layers with two Al compositions ($x \approx 0.06$ or $0.17$). (b) 1D atomic concentration profile of Al, Ga, and Si for Al ≈ 0.17. (c) Corresponding 1D profile for Al ≈ 0.06. Vertical dashed lines indicate the top, mid, and bottom doped layers.

To resolve the spatial origin of this variation, layer-resolved lateral analysis was performed using two-dimensional contour plots (2DCP) of the Al/Ga ratio. As shown in **Figure 2**, the three-dimensional atom map identifies the Si-doped layers, and the corresponding Al/Ga ratio maps reveal pronounced lateral compositional variation exclusively in the top layer with ~$10^{20}$ $cm^{-3}$ Si concentration. In contrast, the middle and bottom layers remain chemically uniform, indicating that moderate Si doping alone is insufficient to induce lateral inhomogeneity.

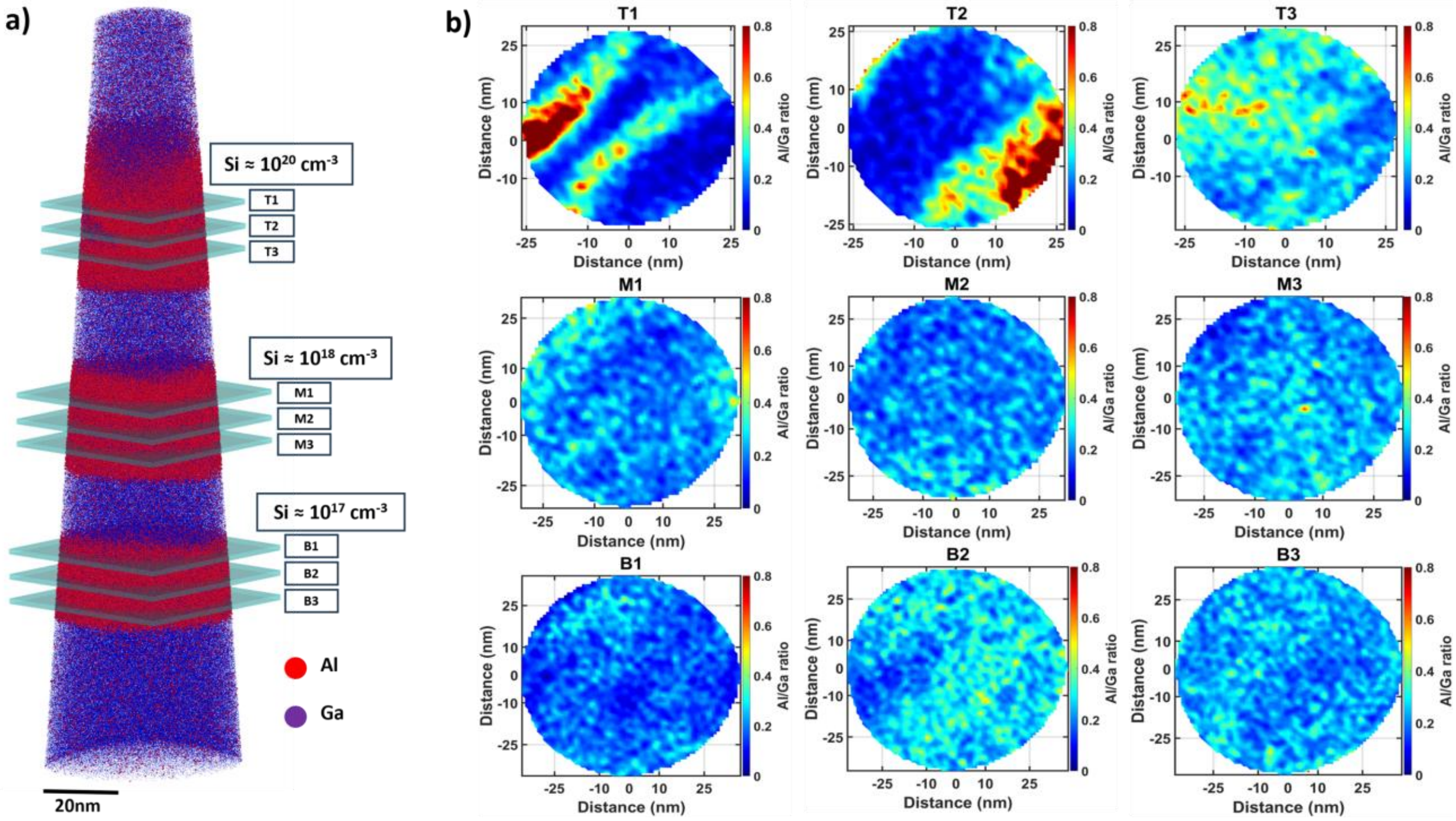

**Figure 2**. (a) 3D atom map of the $(Al_{0.17}Ga_{0.83})_2O_3$ structure showing only Al (red) and Ga (purple) atoms for clarity. Three β-$(Al_{0.17}Ga_{0.83})_2O_3$ layers with varying Si doping concentrations (~$10^{20}$, $10^{18}$, and $10^{17}$ $cm^{-3}$) are indicated as top (T), middle (M), and bottom (B) layers, respectively. (b) Al/Ga ratio maps for selected ROIs (T1-T3, M1-M3, B1-B3) from each layer, showing lateral compositional distributions. Strong Al-rich segregation is observed only in the top Si ≈ $10^{20}$ $cm^{-3}$ layer (T1-T3), while middle and bottom layers show uniform Al/Ga ratios.

To isolate the role of Al composition, the $(Al_{0.06}Ga_{0.94})_2O_3$ heterostructure was analyzed under identical doping conditions (**Figure S2**). Across all layers and doping levels, the Al/Ga distribution remains uniform, with no observable compositional modulation even at the highest Si concentration. Together, these observations establish that lateral compositional variation and the associated structural instability is specific to the combined presence of elevated Al content (~17%) and high Si doping (~$10^{20}$ $cm^{-3}$). Since all layers were grown under identical conditions, this behavior cannot be attributed to growth parameters but instead reflects a composition- and doping-driven instability within the cation sublattice.

To identify the structural origin of the observed compositional variation, high-resolution scanning transmission electron microscopy combined with EDX mapping was performed on the $(Al_{0.17}Ga_{0.83})_2O_3$ heterostructure (**Figure 3**). Distinct γ-phase inclusions, characterized by contrast variation and lattice distortion, are observed exclusively in the top Si ≈ $10^{20}$ $cm^{-3}$ layer. These regions spatially coincide with Al-rich areas identified in the APT analysis. The observed γ-phase features are consistent with previously reported high-Al $(Al_XGa_{1-X})_2O_3$ systems (Al > 40%).[22] Their presence at a substantially lower Al concentration (~17%) indicates that heavy Si doping plays a critical role in promoting γ-phase formation. In contrast, no such features are observed in the lower-doped layers, consistent with the absence of compositional variation in APT. The spatial correlation between Al-rich regions and γ-phase inclusions establishes that the observed lateral compositional variation corresponds to localized structural transformation, confined to heavily doped regions of the heterostructure.

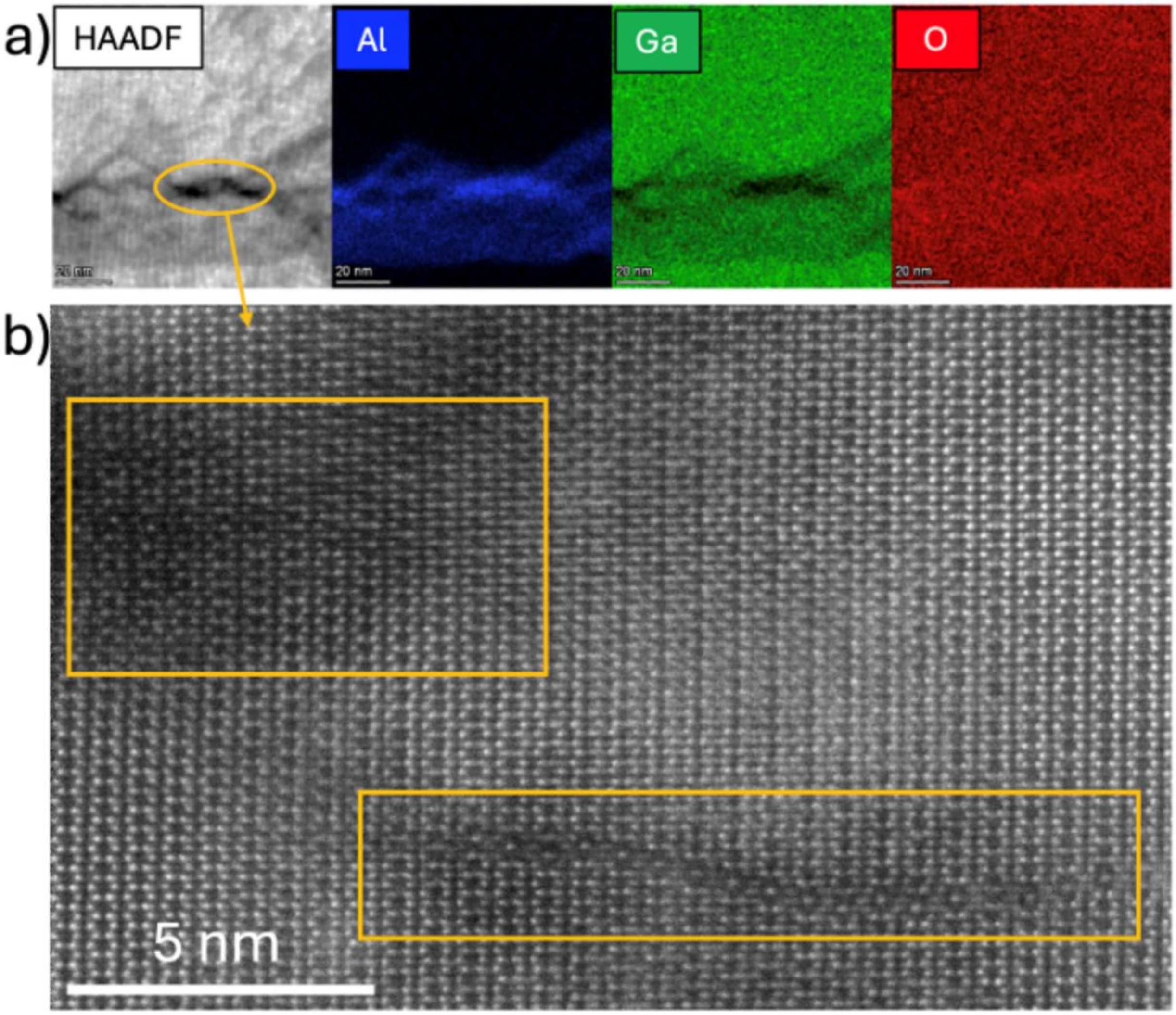

**Figure 3.** (a) STEM HAADF image and corresponding EDX elemental maps (Al, Ga, and O) of the top $(Al_{0.17}Ga_{0.83})_2O_3$ layer with a Si concentration of ~$10^{20}$ $cm^{-3}$. (b) High-resolution HAADF image of the β-$(Al_{0.17}Ga_{0.83})_2O_3$ layer grown along the (010) direction, showing localized contrast variations associated with γ-phase inclusions (yellow rectangles).

The lateral compositional variation observed in the $(Al_{0.17}Ga_{0.83})_2O_3$ layer with ~$10^{20}$ $cm^{-3}$ Si concentration, together with the presence of γ-phase inclusions in the same region, suggests that the chemical inhomogeneity reflects an underlying structural transformation. To evaluate this, local cation coordination environments were analyzed using Ga–Ga first nearest-neighbor (1NN) distributions and Ga-centered radial distribution functions (RDF). A representative of β-phase region is shown in **Figure 4a** region of interest **(ROI) (‘B’)**. The Ga–Ga 1NN distribution exhibits a narrow, near-symmetric profile centered at ~0.3 nm, corresponding to the expected first coordination shell in β-$Ga_2O_3$, and indicative of an ordered local coordination environment. The corresponding Al/Ga map confirms that this ROI lies within a chemically uniform region. The RDF displays a well-defined first-shell feature, and the summed Ga counts within 0.3 nm remains high, consistent with preserved local coordination in the β phase. Together, these observations indicate a structurally ordered region with intact cation coordination. In contrast, **Figure 4b** shows a representative region **(ROI 'G')** of γ -phase extracted from an Al-rich, compositionally inhomogeneous zone. Here, the Ga–Ga 1NN distribution becomes broader and asymmetric, indicating disruption of the local coordination environment. The corresponding RDF exhibits a decreased first-shell feature, and the summed Ga counts within 0.3 nm are substantially reduced relative to the β-phase region. These changes indicate a depletion of Ga neighbors and a locally disordered cation environment in the γ-associated regions.

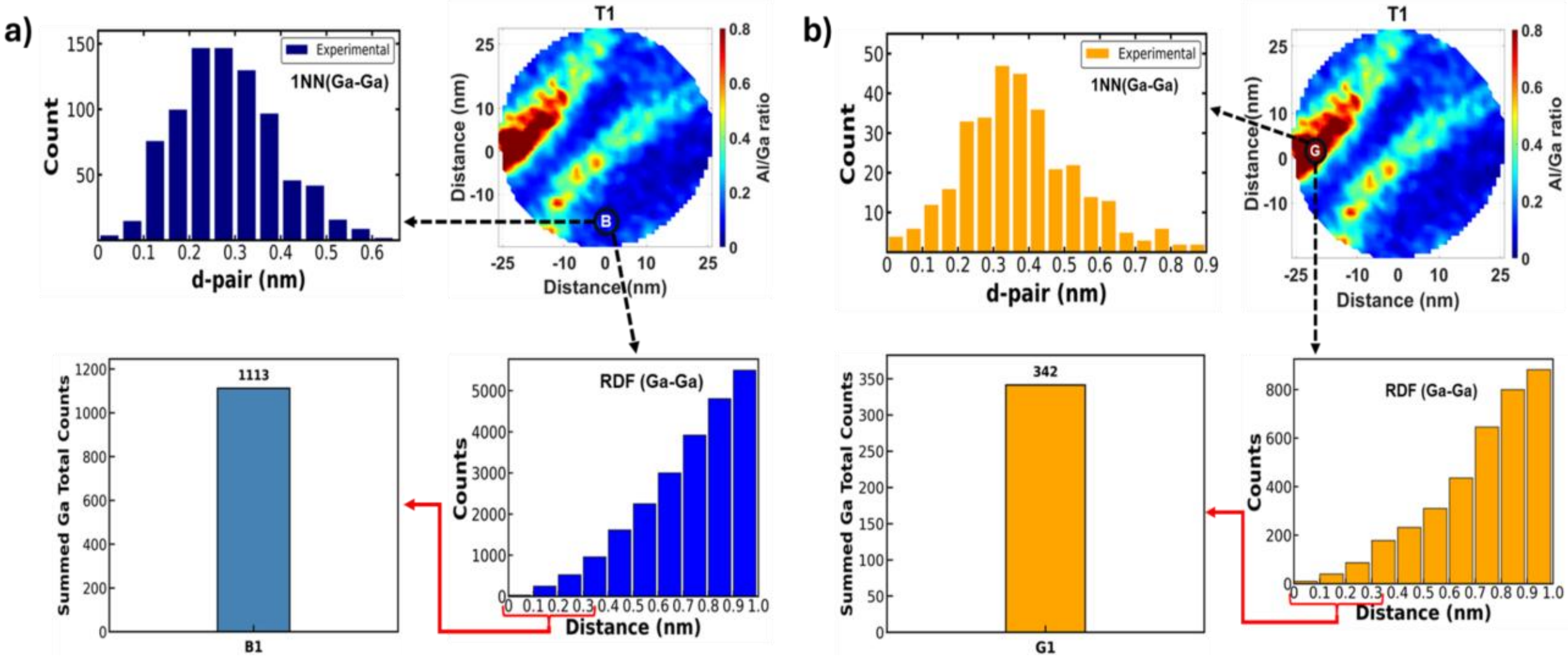

**Figure 4.** Local analysis of β- and γ-phase regions extracted from the same Si ≈ $10^{20}$ $cm^{-3}$, Al ≈ 0.17 layer. (a) β-phase region (B): (top-left) Ga-Ga 1NN distribution showing a near-bell-shaped profile centered at ~0.3 nm, (top-right) 2D Al/Ga ratio map with the ROI location marked, (bottom-left) integrated Ga total ion counts within 0.3 nm from RDF using a 0.05 nm bin width, and (bottom-right) RDF (total Ga-Ga counts) showing well-defined shell structure. (b) γ-phase region (G): (top-left) Ga-Ga 1NN distribution with a broader, slightly asymmetric profile, (top-right) 2D Al/Ga ratio map showing ROI location, (bottom-left) integrated Ga total ion counts within 0.3 nm from RDF using a 0.05 nm bin width, and (bottom-right) RDF curve with reduced integrated Ga counts and less sharply defined shell structure indicative of Ga-vacancy complex formation and local coordination disruption.

To quantify this coordination, change more rigorously, Ga-centered RDFs were evaluated using bin widths of 0.01, 0.02, and 0.05 nm. Because peak position and intensity in APT-derived RDFs can be influenced by trajectory aberrations, reconstruction uncertainty, and local lattice distortion, a single peak metric may not reliably represent first-shell coordination. Therefore, first-shell coordination was assessed using the summed Ga counts within a cutoff of 0.3 nm. This approach captures Ga neighbors that may be displaced from ideal lattice positions while preserving sensitivity to coordination loss within the first shell. Across multiple ROIs, β-phase regions consistently exhibit higher first-shell summed Ga counts, whereas γ-associated ROIs show substantially lower values, indicating significant depletion of Ga neighbors in the transformed regions. The robustness of this trend across multiple ROIs with 0.05 nm RDF bin width is provided in **Figure S1** in the supporting information. Ten γ-phase ROIs and ten β-phase ROIs were evaluated from the top (T1) layer of the $(Al_{0.17}Ga_{0.83})_2O_3$ heterostructure (Si ≈ $10^{20}$ $cm^{-3}$), and a clear separation between the two populations is retained across all binning conditions. The ROI selection and corresponding summed Ga counts are provided in **Figure S1**, which includes the complete dataset used for this analysis. Taken together, these features are consistent with previous experimental observations and theoretical predictions indicating that vacancy formation and atomic migration, particularly involving Ga, distort the local coordination shell and promote structural destabilization.[22,30,52] To assess the statistical significance of the observed coordination difference, two-sample t-tests were performed for each bin size using the mean first-shell summed Ga counts of the β- and γ-associated ROI groups. The results are summarized in **Table S1**, where β-phase regions exhibit higher mean first-shell Ga coordination than γ-associated regions across all binning conditions. This separation is reflected in large t-statistics and negligible p-values, confirming that the observed coordination depletion is statistically significant and not attributable to sampling variability or reconstruction-related effects. These results further support a systematic reduction in first-shell Ga coordination in γ-associated regions, consistent with local cation deficiency and structural disorder.

The observed reduction in first-shell Ga coordination and the associated structural disorder is consistent with a vacancy-mediated mechanism for γ-phase formation. Prior experimental and theoretical studies have shown that heavy donor doping in β-$Ga_2O_3$ leads to the formation of compensating Ga vacancies, which act as dominant defects in n-type material and contribute to local cation deficiency[18,30,52]. These vacancy-rich environments are known to distort the β-phase lattice and promote the formation of disordered spinel-like motifs characteristic of the γ-phase. First-principles studies further indicate that increasing Ga vacancy concentration reduces the energetic stability of the β-phase relative to the γ-phase, facilitating the transformation under conditions of sufficient defect accumulation and thermal activation.[52] In combination with alloying, these effects are further amplified, as Al incorporation modifies local strain and site occupancy, enhancing the susceptibility of the cation sublattice to disorder. In this context, the simultaneous presence of high Si doping and elevated Al content provides a favorable condition for vacancy formation and structural reorganization, consistent with the experimentally observed phase instability in the $(Al_{0.17}Ga_{0.83})_2O_3$ layer.

This interpretation is further supported by prior studies of β-$(Al_XGa_{1-X})_2O_3$ and $Ga_2O_3$ systems. Thermodynamic analysis of the γ-phase in unalloyed and Al-alloyed $Ga_2O_3$ have consistently shown that these phases are higher in energy when small model supercells to represent the disordered sublattice, even when configurational entropy is taken into account.[53] However, Bhuiyan et al. reported γ-phase formation at high Al concentrations (27–40%), attributing the transition primarily to strain effects.[54] Subsequent studies have shown that γ-phase inclusions can emerge at lower Al content when defect accumulation and dopant incorporation are significant.[27] In particular, Huang et al. demonstrated that high Si concentrations can induce defect clustering and local lattice disruption, leading to γ-phase nucleation at concentrations approaching ~$10^{20}$ $cm^{-3}$.[41,55] These findings are consistent with the present results, where γ-phase inclusions are observed only in the highest doped $(Al_{0.17}Ga_{0.83})_2O_3$ layer and are accompanied by measurable Ga coordination loss. More broadly, these results support a unified picture in which Ga-vacancy-driven coordination loss, amplified by dopant incorporation and alloy composition, governs the β-to-γ transformation. In the present system, the combination of heavy Si doping and elevated Al content enhances local defect accumulation and promotes phase destabilization, providing a direct nanoscale pathway for γ-phase formation. Although APT does not directly resolve individual vacancies, the combined coordination-sensitive analyses, local compositional depletion, and prior theoretical predictions collectively provide strong evidence consistent with vacancy-mediated cation redistribution.

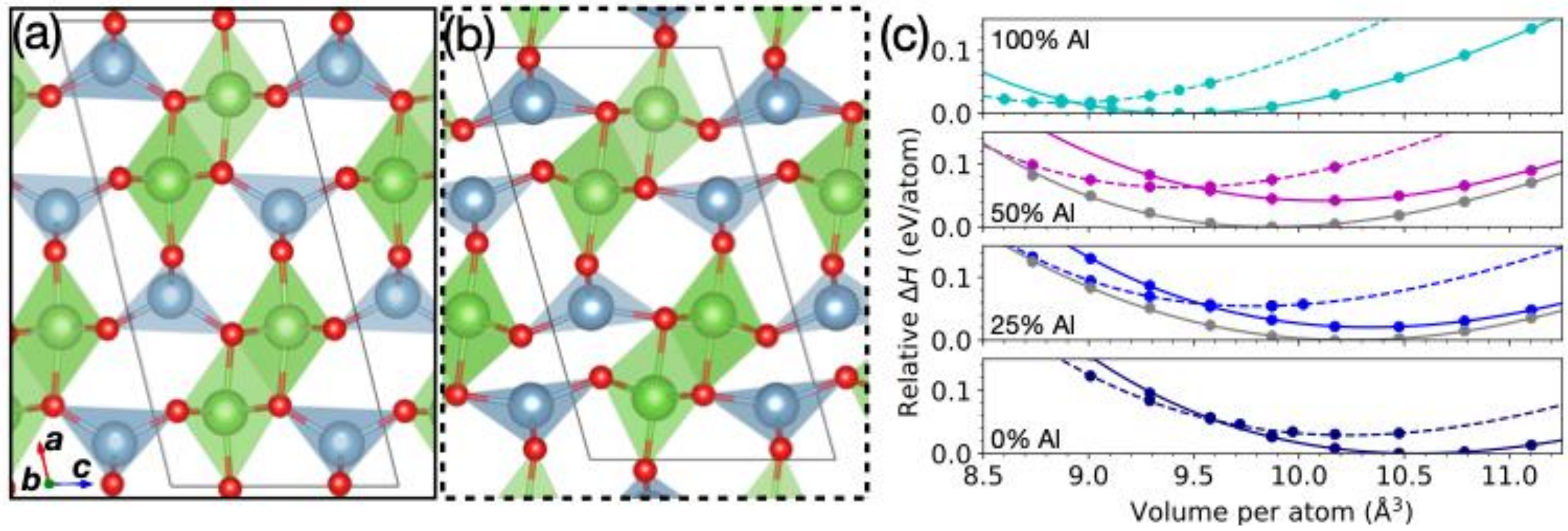

**Figure 5.** Representations of the (a) β-$Ga_2O_3$ and (b) variant HDM monoclinic structures, where the tetrahedral sites are colored blue, and the octahedral sites colored green. Both structures exhibit similar *b*-axis lattice parameters, but differ in the *a*, *c* and *β* parameters. (c) Calculated formation energies for the two monoclinic variants as a function of the volume for different Al contents, with the conventional β phase shown with solid lines and the HDM variant with dashed lines. In (c) the formation enthalpies per atom for each phase with respect to the lowest energy monoclinic structure of each composition. For the alloys in (c), the lowest energy configurations where Al incorporates on the octahedral sites are shaded grey.

To further elucidate the atomistic mechanisms underlying this phase transformation and assess the role of defect-driven structural reorganization, we performed density functional theory (DFT) simulations to investigate both the thermodynamic and kinetic aspects of the β-to-γ transformation. As shown in **Figure 5,** monoclinic $(Al_XGa_{1-X})_2O_3$ exhibits a relatively shallow potential energy landscape that is sensitive to local volume, cation identity and site occupation, including tetrahedral vs octahedral coordination. To capture this structural flexibility, we compare the conventional β-$Ga_2O_3$ structure with a denser monoclinic variant, referred to here as the high-density monoclinic (HDM) structure. These two monoclinic configurations accommodate different combinations of *a, c* and *β* lattice parameters while maintaining relatively similar *b*-axis parameters, as summarized in **Figure S4**. The relative energy difference between the conventional β and HDM structures decreases systematically with increasing Al content, from 30 meV/atom for $Ga_2O_3$ to 21 meV/atom for $(Al_XGa_{1-X})_2O_3$ with tetrahedral Al occupation, and further to 17 meV/atom for $Al_2O_3$. In contrast, when Al exclusively occupies octahedral sites, the HDM configuration relaxes back toward the conventional β-phase. These trends indicate that Al incorporation reduces the energetic separation between competing monoclinic configurations and increases the susceptibility of the lattice to local structural distortion, highlighting the critical role of cation site occupation in stabilizing the monoclinic lattice.

The elastic trends summarized in **Table 1** reveal that tetrahedral-site occupation by Al strongly modifies the mechanical response of the monoclinic lattice. In pure $Ga_2O_3$, the HDM phase remains mechanically more compliant than the conventional β-phase. However, tetrahedral Al incorporation introduces substantial localized chemical pressure that stabilizes the HDM structure while simultaneously reducing its shear resistance. For the 50% alloy with tetrahedral Al occupation, the HDM phase exhibits increased bulk stiffness (K = 211 GPa versus 186 GPa for the β-phase) accompanied by a pronounced reduction in shear modulus (G = 69 GPa versus 93 GPa). Based on the associated volume contraction and elastic response of the β phase lattice, we estimate an effective chemical pressure of $\Delta P$ = ~23 GPa for this case from the expression $\Delta P = K_\beta\ \Delta V/V_\beta$. These results indicate that tetrahedral Al occupation generates significant chemomechanical frustration within the monoclinic lattice, producing structural instability that precedes transformation toward the disordered γ phase.

**Table 1**. Calculated elastic moduli, bulk $K$ and shear $G$, longitudinal stiffness constants ($C_{ii}$), the Pugh ratio ($K/G$), the elastic anisotropy index ($A^U$) and the Cauchy pressure (defined as $C_{12}$-$C_{44}$), summarized for $Ga_2O_3$, $Al_2O_3$, and ordered $AlGaO_3$ alloys where Al occupies different sites (tetrahedral vs octahedral) in the HDM and β phases.

| **Material** | **Phase (site)** | ***K* (GPa)** | ***G* (GPa)** | **$C_{11}$ (GPa)** | **$C_{22}$ (GPa)** | **$C_{33}$ (GPa)** | ***K/G*** | **$A^U$** | **Cauchy pressure (GPa)** |
|---|---|---|---|---|---|---|---|---|---|
| $Ga_2O_3$ | β | 173 | 81 | 218 | 337 | 320 | 2.1 | 0.9 | 78.1 |
| $Ga_2O_3$ | HDM | 150 | 71 | 137 | 291 | 306 | 2.1 | 2.1 | 11.8 |
| $AlGaO_3$ | β (Oct) | 192 | 98 | 253 | 374 | 372 | 2.0 | 0.8 | 66.1 |
| $AlGaO_3$ | β (Tet) | 186 | 93 | 244 | 378 | 349 | 2.0 | 0.7 | 75.6 |
| $AlGaO_3$ | HDM (Tet) | 211 | 69 | 344 | 393 | 330 | 3.1 | –12.8 | 46.5 |
| $Al_2O_3$ | β | 197 | 111 | 262 | 401 | 398 | 1.8 | 0.7 | 48.2 |
| $Al_2O_3$ | HDM | 216 | 90 | 349 | 419 | 391 | 2.4 | –6.6 | –48.1 |

Kinetic considerations further support the γ-phase transformation, as compensating cation vacancy acceptors accompanying donor doping are known to reorient at low temperatures and facilitate the transformation toward the γ-phase.[50,52] For example, donor migration in unalloyed $Ga_2O_3$ has been shown to be mediated by cation vacancies, with migration barriers for cation vacancies and interstitials reported to occur at or well below the growth temperatures typical of MOCVD around 800-900 °C.[28,56–58] As discussed previously, cation vacancies concentrations increase with higher incorporated donor concentrations such as Si, and are also expected to further increase with Al incorporation under donor-doped conditions.[18,22,59–61] In pure β-$Ga_2O_3$, higher concentrations of cation vacancies can facilitate a redistribution from tetrahedrally-coordinated to octahedrally-coordinated cation sites ($O_h/T_d$ > 0.5) through the formation of more favorable off-site split-vacancy configurations, analogous to local structural motifs associated with the crystalline γ phase.[28,52,62–66]

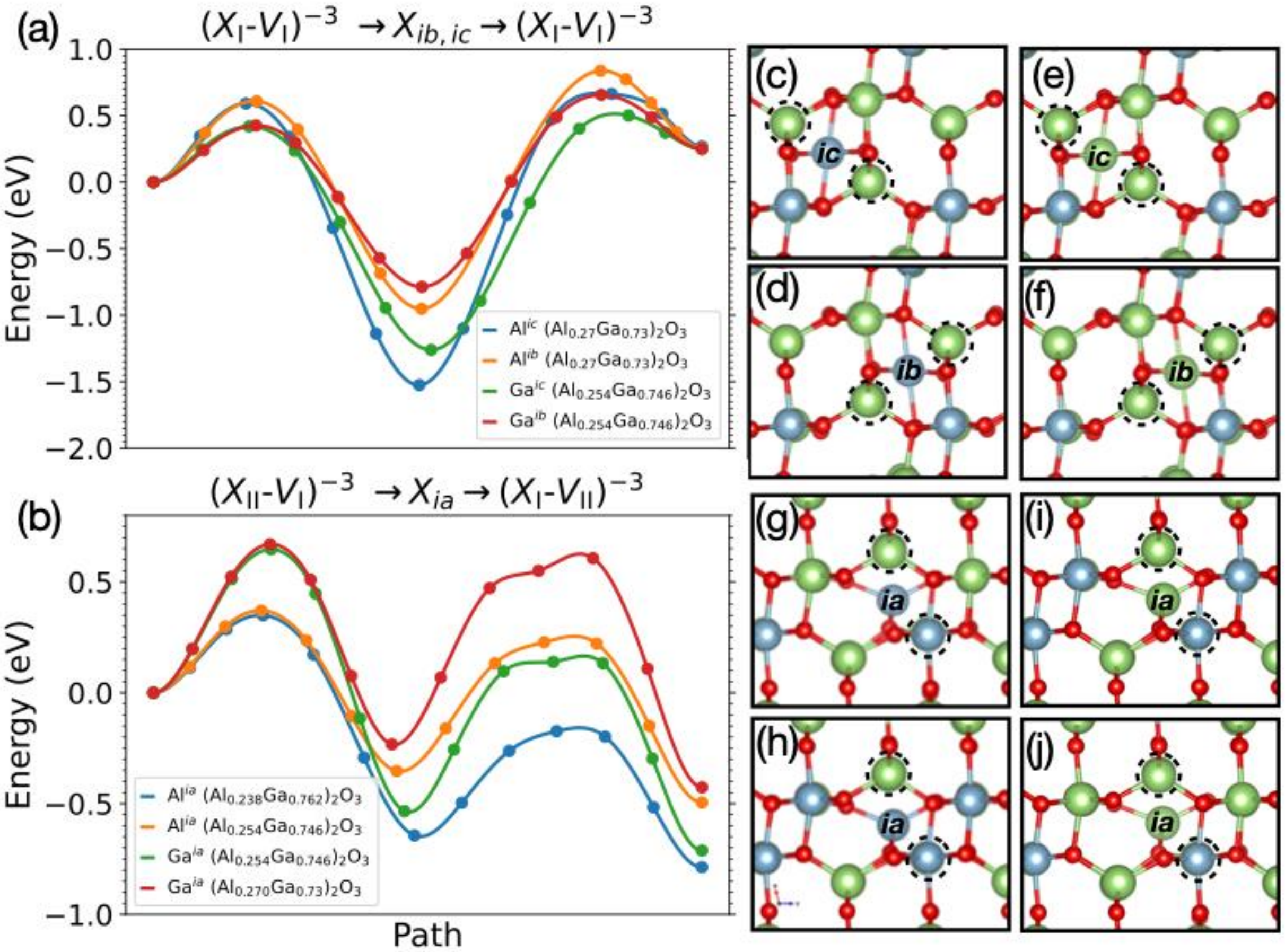

**Figure 6.** Calculated barriers for vacancy-assisted migration of Ga and Al through tetrahedral vacancies via the offsite $V^{\text{ib}}$ and $V^{\text{ic}}$ configurations (a), and the metastable $V^{\text{ia}}$ configuration (b), referenced to the energies of $V_{\text{I}}$ in model $(Al_XGa_{1-X})_2O_3$ alloys representative of the experimental alloy composition. For each supercell missing a single cation, the corresponding composition of Al and Ga are included in parentheses within the legends. Images for the offsite minima in (a) are shown in (c) and (d) for Al (blue atoms) and (e) and (f) for Ga (green atoms), also following the order in the legend. Images for the offsite *ia* minima in (b) are shown in (g) and (h) for Al and (i) and (j) for Ga, following the order in the legend. All barriers are shown for the –3 charge state favorable in *n*-type conditions.

We find that similar vacancy-assisted reorientation mechanisms also occur in β-$(Al_XGa_{1-X})_2O_3$ alloys, where tetrahedral-site vacancies ($V_{\text{I}}$) undergo facile reorientation to the split vacancy configurations under *n*-type conditions, including the $V^{\text{ib}}$ and $V^{\text{ic}}$ split vacancies, or to $V_{\text{II}}$ through the metastable $V^{\text{ia}}$ configurations. To compare more directly with the experimental compositions, we focus on model β-phase alloys containing nominally 25% Al using a 160-atom supercell with one cation site removed, and Al occupying exclusively octahedral sites unless otherwise specified in **Figure 6.** The simulations cover a range of alloy compositions representative of the experimentally relevant mid to high 20% Al regime. As summarized in **Figure 6**, both Al and Ga neighboring $V_I$ exhibit low migration barriers (~0.7 eV) and favorable energy gains for forming $V^{ib}$ and $V^{ic}$ split vacancies (**Figure 6a**), or for exchanging with adjacent octahedral sites to form $V_{II}$. Several important trends emerge from these calculations. First, Al incorporated on tetrahedral sites preferentially gain more energy than Ga when in the proximity of an $V_I$, with the $V^{ic}$ more favorable than the $V^{ib}$ configurations for both cation types, with the offsite energy gains correlating with the proximity of the second nearest neighbor shells (**Figure 6 and Supporting Information**).

From **Figure 6**, we further find that Ga generally moves off-site more readily than Al from either octahedral or tetrahedral sites. For example, the calculated barriers for off-site displacement from octahedral positions to form metastable $V^{ia}$ configurations are ~0.35 eV for Ga compared to ~0.65 eV for Al. The $V^{ia}$ configuration also exhibits less sensitivity to the local environment than displacements originating from tetrahedral sites **(Supporting Information)**. Similar trends are observed for migration from $V^{ia}$ back to tetrahedral or octahedral configurations, where Al consistently exhibits the largest migration barriers, ranging from 0.7–0.85 eV for migration toward octahedral sites and 0.9–1.2 eV for returning to tetrahedral sites. Overall, the migration barriers for Al are 0.26 ± 0.05 eV higher than those of equivalent Ga migration pathways in the configurations studied. These results indicate that vacancy-mediated reorientation of tetrahedrally coordinated Al is kinetically hindered relative to Ga, while elevated populations of tetrahedral Al simultaneously promote chemomechanical frustration within the alloy lattice. Together, these effects point to the combined influence of donor doping and increased Al incorporation in driving γ-phase formation in β-$(Al_XGa_{1-X})_2O_3$.

Together, these results establish coordination-sensitive APT analysis as a quantitative nanoscale approach for probing defect-driven structural instability in complex oxide semiconductors. The combined experimental and theoretical results reveal a unified mechanism in which Al-induced lattice flexibility and vacancy-mediated cation rearrangement collectively drive coordination loss and promote β-to-γ transformation in Si-doped $(Al_XGa_{1-X})_2O_3$. By directly correlating local coordination depletion with compositional inhomogeneity and structural disorder, this work provides a transferable framework for investigating defect-mediated phase transformations in doping- and strain-sensitive ultrawide bandgap materials. More broadly, these findings highlight the critical role of local coordination environments in governing phase stability and structural reliability in complex oxide semiconductors.

## CONCLUSIONS

This work establishes coordination loss as a measurable nanoscale signature of defect-driven phase transformation in Si-doped $(Al_XGa_{1-X})_2O_3$. Using a coordination-sensitive APT-based framework, we show that γ-phase formation is accompanied by a systematic depletion of first-shell Ga coordination, providing strong nanoscale evidence consistent with vacancy-mediated cation deficiency. This instability emerges only under the combined conditions of elevated Al content and heavy Si doping, revealing a synergistic mechanism that destabilizes the β-phase lattice. DFT further supports that Al-induced lattice flexibility, together with donor-driven vacancy formation, promotes cation reconfiguration and structural transformation toward the γ-phase. Beyond this specific system, the ability to directly quantify coordination loss provides a new pathway to identify early-stage structural instability in doped and alloyed materials. This framework is broadly applicable to UWBG semiconductors and complex oxides, where defect-mediated transformations critically impact electronic performance and reliability, enabling more physically informed strategies for defect and phase engineering.

## MATERIALS AND METHODS

### Film Growth and Heterostructure Design

$(Al_XGa_{1-X})_2O_3$ films with Al contents of 6% and 17%, doped with n-type Si concentrations of ~$10^{20}$, $10^{18}$, and $10^{17}$ $cm^{-3}$, were grown by metal organic chemical vapor deposition on Fe-doped semi-insulating (010) β-$Ga_2O_3$ substrates (Novel Crystal Technology, Inc.). Detailed growth conditions are reported elsewhere.[67] The heterostructures consist of three Si-doped layers with varying doping concentrations to enable systematic evaluation of doping-dependent structural evolution. To facilitate atom probe tomography specimen preparation, a ~50 nm unintentionally doped (UID) β-$Ga_2O_3$ sacrificial layer was deposited on top of the $(Al_XGa_{1-X})_2O_3$/β-$Ga_2O_3$ heterostructures. This layer protects the region of interest during focused ion beam (FIB) processing and ensures reliable tip fabrication.

### APT Analysis

APT specimens were prepared using a standard site-specific FIB lift-out technique followed by annular milling.[68] A final low-kV cleaning step was applied to minimize Ga ion implantation and surface damage. APT experiments were conducted using a pulsed laser-assisted CAMECA LEAP 5000 XR system under ultrahigh vacuum ($<10^{-11}$ mbar) at a base temperature of 50 K. Data acquisition was performed using a laser pulse energy of 20 pJ and a pulse repetition rate of 200 kHz, with a detection rate of 0.005 atoms per pulse. These conditions were selected to ensure stable evaporation behavior and to minimize trajectory aberrations and reconstruction artifacts.

Three-dimensional reconstructions were performed using CAMECA Integrated Visualization and Analysis Software (IVAS, version 6.3.1.110). Multiple datasets were collected for each heterostructure to ensure reproducibility. Representative reconstructions are presented in this work to capture the structural and chemical characteristics of the layers.

### Local Coordination Analysis Using 1NN and RDF Metrics from APT Data

To determine whether the observed chemical fluctuations correspond to measurable structural changes, local cation coordination environments were analyzed using two complementary metrics: first nearest-neighbor distributions and radial distribution functions.

First, Ga–Ga 1NN distance distributions were evaluated to quantify the separation between neighboring Ga atoms. This metric provides a direct measure of local ordering, where narrow and symmetric distributions indicate well-defined coordination consistent with the β phase, while broadened or asymmetric distributions reflect distortion of the local cation environment. Second, Ga-centered RDFs were computed to characterize the radial distribution of neighboring Ga atoms. RDFs were evaluated in two complementary forms. In the first, binned Ga–Ga neighbor counts were plotted as a function of radial distance, capturing variations in local neighbor density. In the second, the total first-shell coordination was quantified by summing the Ga counts within a defined cutoff radius. For this purpose, a cutoff of 0.3 nm was used, corresponding to the expected first coordination shell in β-$Ga_2O_3$. While APT does not directly resolve crystallographic coordination, relative changes in first-shell Ga neighbor statistics provide a sensitive comparative metric for distinguishing structurally ordered and coordination-depleted regions. This approach accounts for slight displacements of atoms from ideal lattice positions arising from experimental effects and growth-induced structural distortions, enabling reliable comparison of local coordination environments across different regions.

**STEM Characterization**

Structural characterization was conducted using high-angle annular dark-field scanning transmission electron microscopy. Cross-sectional TEM lamellae were prepared using a dual-beam focused ion beam system (FEI Helios NanoLab 600) employing a conventional lift-out approach. To reduce ion-beam-induced damage and surface amorphization, the specimens were subsequently cleaned by low-energy Ar ion milling (Fischione NanoMill) at acceleration voltages of 900 and 500 V with a milling angle of approximately 10° applied from both sides.

HAADF-STEM imaging was performed on a probe-corrected Thermo Fisher Scientific Titan microscope operated at 300 kV. Images were acquired using probe convergence semi-angles of approximately 20-30 mrad and detector collection semi-angles ranging from ~50 to 400 mrad, enabling Z-contrast imaging sensitive to local compositional fluctuations. These imaging conditions facilitate the observation of lattice distortion and nanoscale structural inhomogeneity. The HAADF-STEM analysis was used to identify γ-phase inclusions through contrast and structural variations and to correlate these features with the compositional heterogeneity observed in atom probe tomography.

**Density Functional Theory Calculations**

DFT calculations were performed using the screened hybrid Heyd–Scuseria–Ernzerhof (HSE) functional with a mixing parameter of 32% and a screening parameter of 0.2 $Å^{-1}$. Projector-augmented wave (PAW) potentials were employed as implemented in the Vienna Ab initio Simulation Package (VASP), where the Ga *3d* electrons were treated explicitly as valence states.[69–71] These parameters have been previously validated for monoclinic $Ga_2O_3$ and $Al_2O_3$, yielding band gaps and lattice parameters in good agreement with experimental values.[18,24,72,73] We used PBEsol calculations to more efficiently sample the potential energy landscape associated with different monoclinic structures as a function of Al content, where a range of volumes were scanned for model 10-atom unit cells using an energy cutoff of 520 eV, a Γ-centered k-point-sampling density of 0.16 $Å^{-1}$ and volume-constrained cell-shape optimization with a force tolerance of 0.01 eV/Å fit with a Birch-Murnaghan equation of state. Elastic constants were evaluated for the optimal volume structures for each monoclinic variant using the atomate2 elasticity workflow with PBEsol.[74] From the computed elastic tensor elements $C_{ij}$ and the bulk moduli and shear moduli ($K$ and $G$) we specifically consider key elasticity metrics such as the Pugh ratio ($K/G$), the Cauchy pressure (defined as $C_{12}$-$C_{44}$ for quasi-cubic crystals), and the Universal Anisotropy index ($A^U$), which we describe more in the Supporting Information.[75–77]

A 160-atom supercell was used to model monoclinic $Ga_2O_3$, and $(Al_XGa_{1-X})_2O_3$ alloys were constructed using ordered configurations with 25% and 50% Al occupying all octahedral or tetrahedral sites, along with $Al_2O_3$ for comparison. Vacancy formation energies were evaluated for both onsite and offsite (split-vacancy) configurations following the methodology described previously.[18] Migration barriers between onsite and offsite configurations were calculated using the climbing image nudged elastic band method with five images and a force convergence criterion of 0.03 eV/Å.[78]

**Author Contributions**

S.D. conceptualized the study, performed APT data analysis, and prepared the original manuscript draft. J.S. conducted the APT experiments. J.B.V. performed DFT calculation and analysis. C.C. conducted STEM experiments and analysis. L.M. performed the MOCVD growth of the heterostructures. J.H. supervised the STEM characterization and analysis. H.Z. supervised the material growth. B.M. conceptualized and supervised the overall project and revised the manuscript.

**Notes**

The authors declare no competing financial interest.

**Acknowledgements:**

S.D., J.S. and B.M. would like to acknowledge the support from National Science Foundation (NSF) under the Division of Materials Research (DMR) through Grant No. 2145091. The work of J.B.V. was performed under the auspices of the U.S. Department of Energy by Lawrence Livermore National Laboratory under contract No. DE-AC52-07NA27344. C.C. and J.H. acknowledge support by the Department of Defense, Air Force Office of Scientific Research GAME MURI Program (Grant No. FA9550-18-1-0479). L.M. and H.Z. acknowledge funding support from the Air Force Office of Scientific Research FA9550-18-1-0479 (AFOSR, Dr. Ali Sayir), and the National Science Foundation (Grant No. 1810041, No. 2019753).

# Supporting Information

# Coordination-Sensitive Nanoscale Analysis of Defect-Driven Phase Transformation in Si-Doped $(Al_XGa_{1-X})_2O_3$

Shaon Das[1], Jith Sarker[1], Christopher Chae[3], Lingyu Meng[2], Joel B. Varley[4], Hongping Zhao[2, 3], Jinwoo Hwang[3], and Baishakhi Mazumder[1*]

*[1] Department of Materials Design and Innovation, University at Buffalo-SUNY, Buffalo, NY 14260, USA*
*[2] Department of Electrical and Computer Engineering, The Ohio State University, Columbus, OH 43210, USA*
*[3] Department of Materials Science and Engineering, The Ohio State University, Columbus, OH 43210, USA*
*[4] Materials Science Division, Lawrence Livermore National Laboratory, Livermore, CA, USA*

** Corresponding author: baishakh@buffalo.edu*

To determine whether the lateral compositional variation observed in the high Al-content heterostructure corresponds to changes in local cation coordination, analysis was focused on the top (T1) layer of the $(Al_{0.17}Ga_{0.83})_2O_3$ heterostructure (Si ≈ $10^{20}$cm$^{-3}$), where the strongest compositional variation and γ-associated features are observed. Regions of interest (ROIs) were selected from areas exhibiting distinct compositional contrast, and Ga-centered radial distribution function (RDF) analysis was used to quantify first-shell coordination.

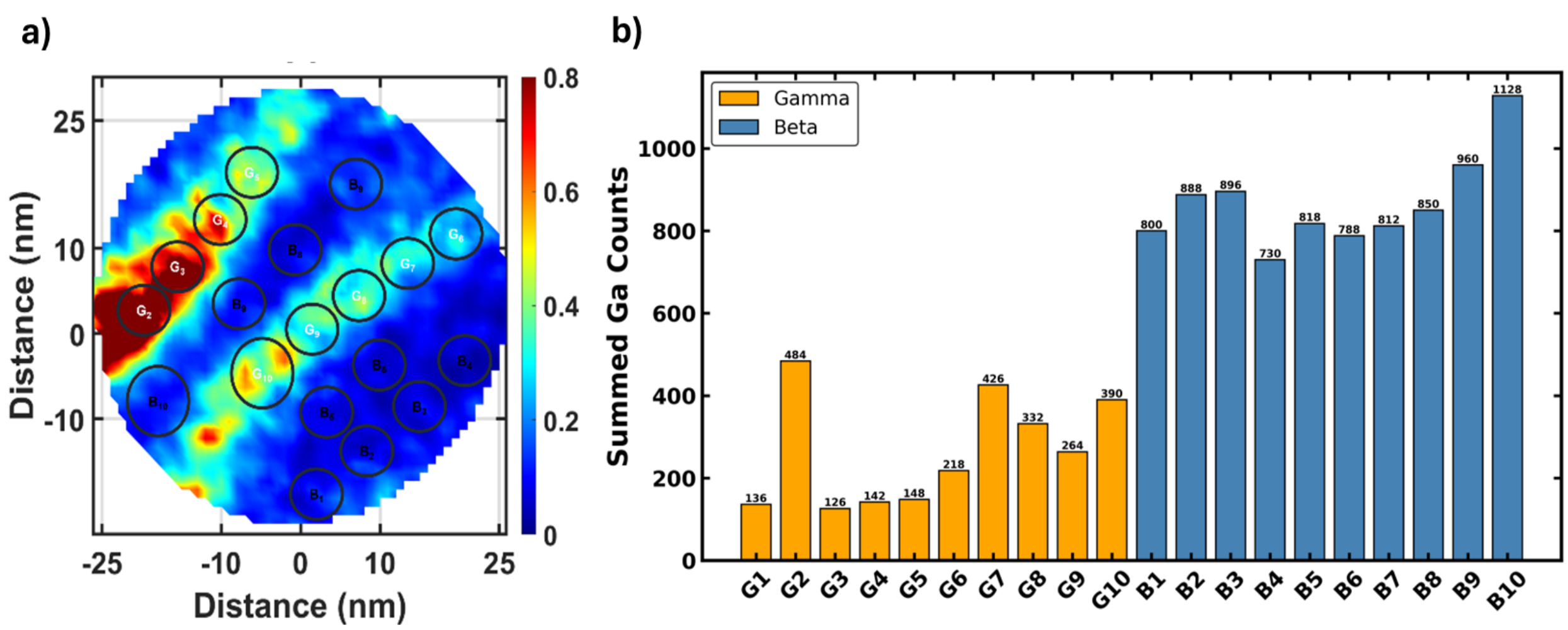

***Figure S1.*** *(a) Al/Ga ratio map from the top (T1) layer of the $(Al_{0.17}Ga_{0.83})_2O_3$ heterostructure (Si ≈ $10^{20}$ cm$^{-3}$), showing lateral compositional variation. Ten regions of interest (ROIs) were selected*

*from γ-associated regions ($G_1$–$G_{10}$) and ten from β-phase regions ($B_1$–$B_{10}$), as indicated by compositional contrast. (b) First-shell Ga coordination for each ROI, quantified as the summed Ga counts obtained from RDF analysis within a cutoff radius of 0.3 nm (bin width = 0.05 nm). ROIs associated with γ-contrast regions exhibit consistently lower Ga counts compared to β-phase regions, indicating reduced first-shell Ga coordination in γ-associated regions.*

To establish a compositional baseline, the $(Al_{0.06}Ga_{0.94})_2O_3$ heterostructure was analyzed across layers with different Si doping concentrations. Three regions corresponding to high (~$10^{20}$ cm$^{-3}$), intermediate (~$10^{18}$ cm$^{-3}$), and low ((~$10^{17}$ cm$^{-3}$) Si doping levels were examined to evaluate the presence of lateral compositional variation in the absence of high Al content.

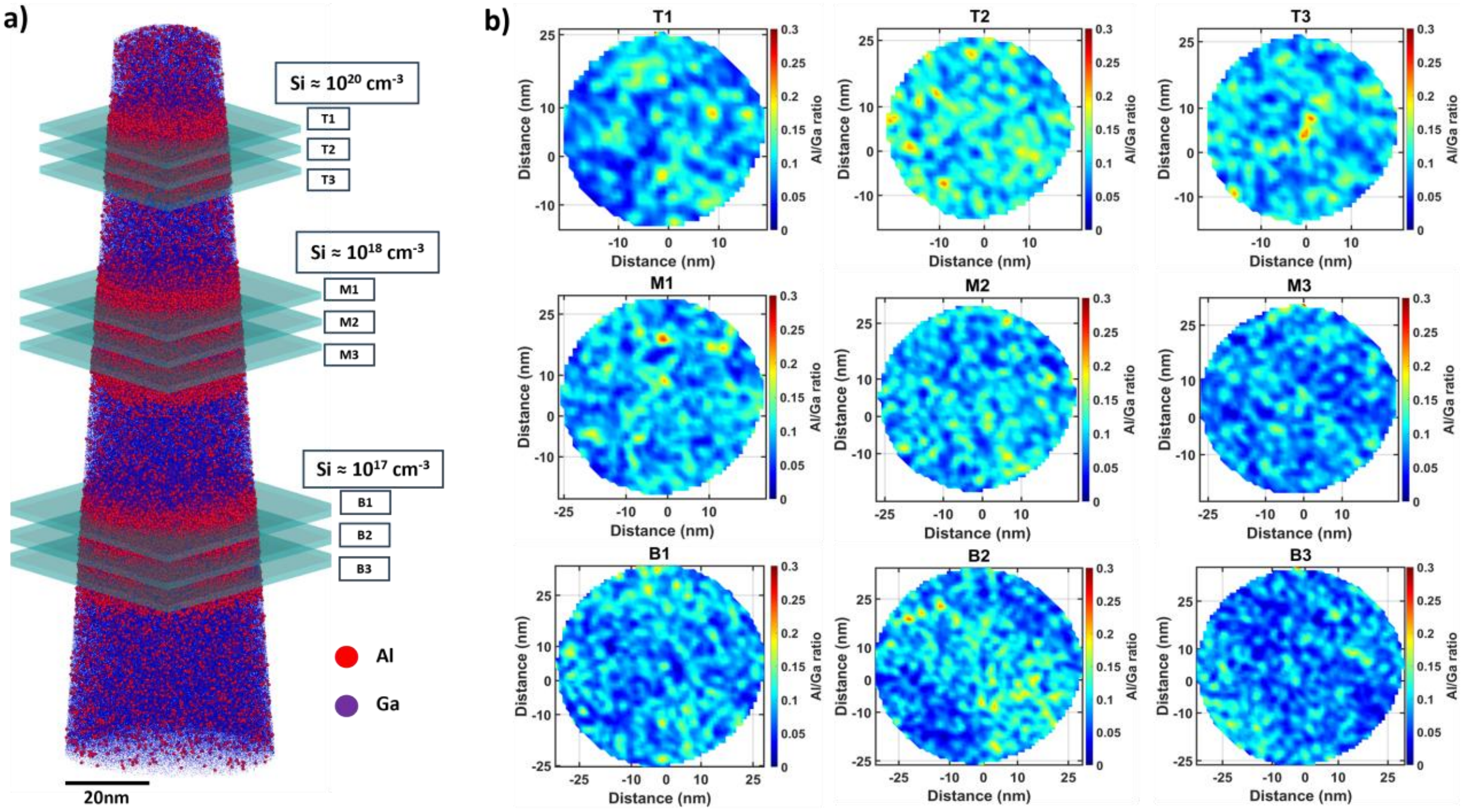

***Figure S2.*** *(a) Three-dimensional atom probe tomography (APT) reconstruction of the $(Al_{0.06}Ga_{0.94})_2O_3$ structure, showing only Al (red) and Ga (purple) atoms. Three β-$(Al_{0.06}Ga_{0.94})_2O_3$ layers with different Si doping concentrations (~$10^{20}$, $10^{18}$, and $10^{17}$ cm$^{-3}$) are identified as top (T), middle (M), and bottom (B) regions, respectively. (b) Spatial maps of the Al/Ga ratio obtained from selected regions of interest (ROIs) within each layer (T1–T3, M1–M3, B1–B3). The Al/Ga distribution remains laterally uniform across all layers, with no discernible compositional variation, even at the highest Si doping level.*

## Normalization of First-Shell Cation Coordination

To compare local coordination environments across regions, the ratio Ga/(Al + Ga) was used instead of absolute Ga counts. This normalization accounts for the presence of both Ga and Al on the cation sublattice and enables comparison of the relative Ga occupancy within the first coordination shell independent of local compositional variations. In the 6% Al heterostructure, the first-shell cation environment remains strongly Ga-dominated across all examined ROIs, with only minor variation in Ga/(Al + Ga) and no shift toward reduced Ga fraction. Although some variability in total first-shell coordination may be present, the absence of correlated Ga depletion, relative Al enrichment, or lateral compositional variation indicates that these regions remain β-like.

To evaluate whether variations in first-shell coordination arise solely from compositional differences or reflect changes in local cation environments, normalized first-shell coordination was analyzed in the $(Al_{0.06}Ga_{0.94})_2O_3$ heterostructure. Regions of interest (ROIs) were selected from the top (T2), middle (M2), and bottom (B2) layers, and first-shell coordination was expressed as Ga/(Al + Ga) to enable comparison of relative Ga occupancy across different regions.

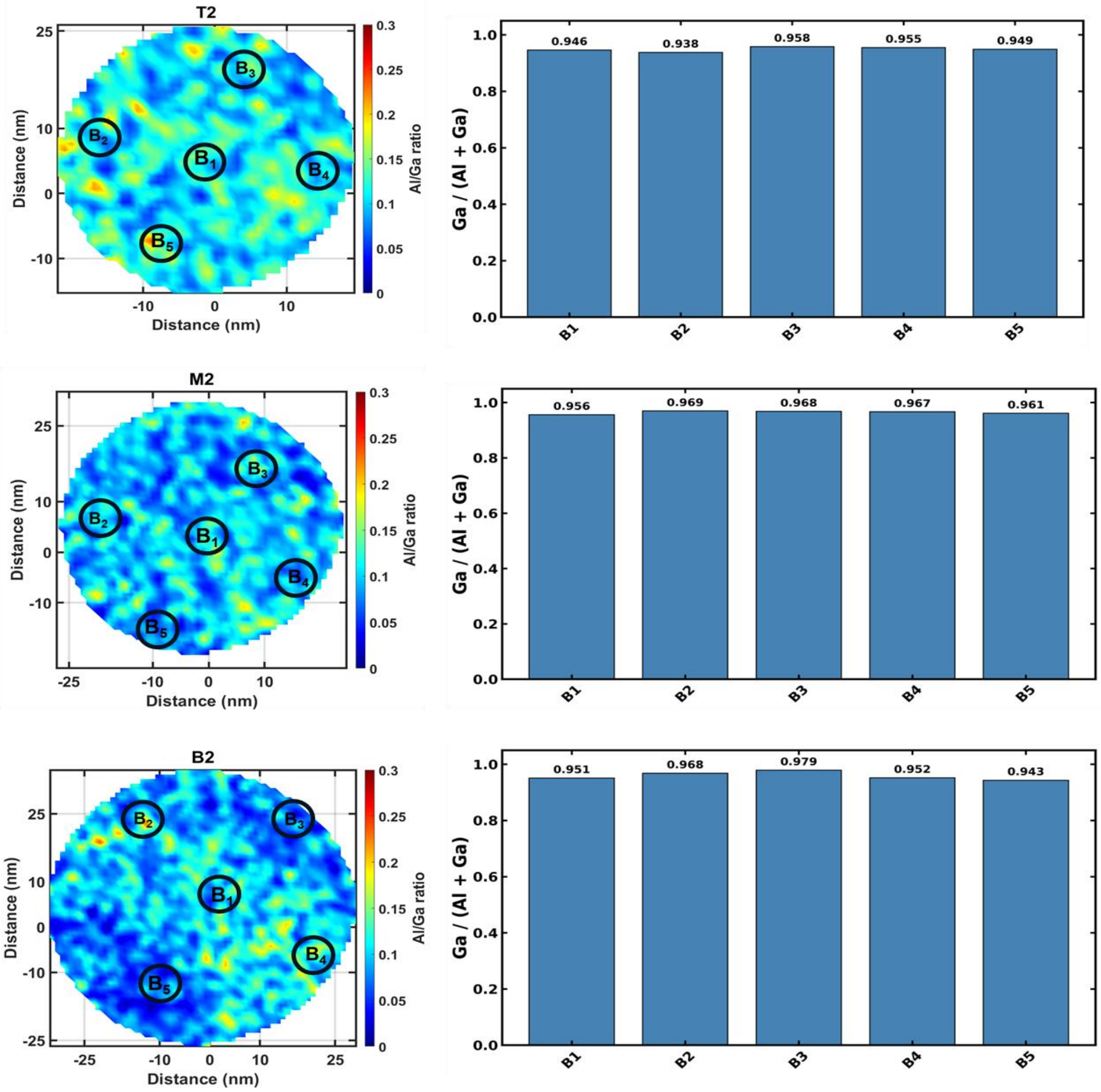

***Figure S3.*** *Al/Ga ratio maps from representative regions of interest (ROIs) in the top (T2), middle (M2), and bottom (B2) layers of the $(Al_{0.06}Ga_{0.94})_2O_3$ heterostructure, along with corresponding first-shell cation coordination expressed as Ga/(Al + Ga) for each ROI. The Ga/(Al + Ga) values are calculated from RDF analysis by summing first-shell counts (≤ 0.3 nm) for Ga and total cations (Al + Ga). Across all ROIs and layers, the cation environment remains strongly Ga-dominated, with only minor variation in Ga/(Al + Ga) and no change in relative cation distribution.*

**Table S1:** Statistical comparison of integrated Ga coordination (within 0.3 nm) between γ-phase and β-phase ROIs at multiple RDF bin sizes from the top (T1) layer of the $(Al_{0.17}Ga_{0.83})_2O_3$ heterostructure (Si ≈ $10^{20}$ $cm^{-3}$).

| Bin Size | Metric | Gamma Mean | Beta Mean | T statistic | P value |
|---|---|---|---|---|---|
| 0.01 | Summed Ga | 289.6 | 951.8 | 11.11 | ~0.0 |
| 0.02 | Summed Ga | 316.8 | 1030.8 | 10.92 | ~0.0 |
| 0.05 | Summed Ga | 266.6 | 867 | 10.88 | ~0.0 |

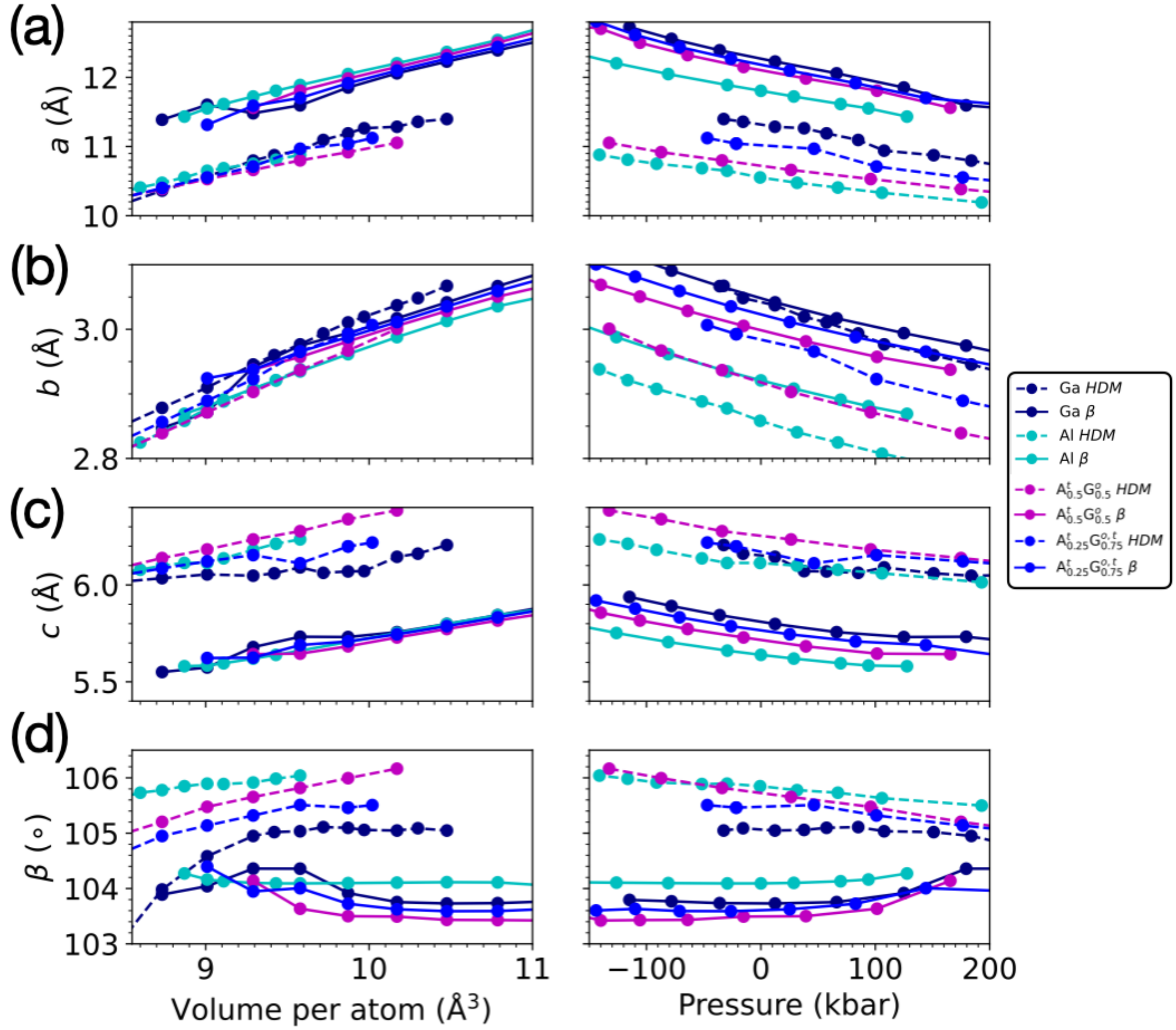

***Figure S4**. Summary of calculated lattice parameters for the conventional β-phase (solid lines) and the HDM variant (dashed lines) for different Al site occupations, shown as a function of the*

*volume per atom (first column) and the external pressure (second column). Ordered alloys for Ga occupying octahedral sites ($Ga^o$) and 25% Al and 50% Al occupying tetrahedral site ($Al^t$) (e.g. half and all of the tetrahedral cation sites within the lattice) are included. The a and c lattice parameters and angle β are found to strongly differ between the HDM and β-phase, while the b lattice parameter is found to be very similar.*

**Physical Significance of Elasticity Criteria**

We evaluated three additional elastic criteria calculated with PBEsol, namely the Cauchy pressure, Pugh ratio, and Universal Anisotropy index, that all point to increasing mechanical instability of the higher-density monoclinic (HDM) variants of $(Al_XGa_{1-X})_2O_3$ that can form with increasing Al incorporation on tetrahedral sites. The Cauchy pressure, which is generally defined as the difference between the Bulk and Shear moduli, provides insight into the nature of the atomic bonding. A positive value indicates that the bonding is primarily non-directional, e.g. ionic or metallic, and correlates with more ductile material that is prone for plastic deformation.[1] For the pseudo-cubic case relevant for the γ-phase, the Cauchy pressure is defined by difference in the elastic constants $C_{12}$-$C_{44}$, and which results in a large, positive Cauchy pressure (+46.5 GPa) in the HDM phase. This suggests a weakening of the directional bonds of the conventional β-phase that facilitate easier atomic reconfiguration preceding γ-phase formation. The Pugh ratio is a measure of the competition between resistance to volume change (bulk modulus $K$) and resistance to shape change (shear modulus $G$), with a critical threshold typically $K/G < 1.75$.[2] Below this value, directional bonding dominates and mechanically brittle behavior, while above this value there is more mechanically ductile behavior. We calculate a Pugh ratio of 3.1 in the HDM phase, which implies a structure that is shear soft while compressively hard and that which will structurally collapse under shear stress rather than cracking.

Within elasticity theory, the universal anisotropy index ($A^U$) was defined by Rangathan and Ostoja-Starzewski as $A^U = 5\ G_V/G_R + 5K_V/K_R - 6$, for the shear and bulk moduli as defined within the Reuss ($R$) and Voigt ($V$) bounds under the assumption of uniform stress and strain, respectively.[3] $A^U$ serves as a scalar measure of the divergence between the Voigt (upper) and Reuss (lower) bounds of the elastic moduli. For a perfectly isotropic crystal, $A^U = 0$, while positive values typically represent the degree of directional variance in standard stable crystals, and negative values imply elastic instability. For the HDM phase, at intermediate Al contents we observe a transition to a large negative value ($A^U = -12.8$ for the 50% Al alloy where Al occupies the tetrahedral sites), which implies the elastic stability limits are exceeded within the monoclinic framework depending on the concentration and site occupation of Al. For full Al occupation, we also observe a negative $A^U$. This suggests that the HDM variant, while compressively stiff as indicated by the higher bulk modulus ($K$ = 211 GPa for 50% Al), possesses specific shear planes with nearly vanishing resistance. Such extreme anisotropy is a known precursor to structural phase transitions or the onset of long-range disorder, as the lattice can no longer uniformly distribute internal stresses. In the context of $(Al_XGa_{1-X})_2O_3$ alloys, this indicates that the HDM structure is a transient, frustrated state that can be resolved through relaxation into another configuration, such as the closely-related γ-phase. Thus overall, the combination of the high Cauchy pressure, high Pugh ratio, and large negative Universal Anisotropy index all support that the HDM phase acts as a mechanical precursor to the structural collapse and long-range disordering characteristic of the transition to the γ-phase transition.